\documentclass{comnet}
\usepackage{amssymb}
\usepackage{amsfonts}
\usepackage{amsmath}
\usepackage{graphicx}%
\usepackage{float}

\begin{document}

\title{Corrigendum to ``Degree-Based Approximations for Network Reliability Polynomials''.
Comment on J. Complex Networks 2025, 13, cnaf001}

\shorttitle{Corrigendum} 
\shortauthorlist{P. Van Mieghem and X. Liu} 

\author{%
  \name{Xinhan Liu \thanks{Corresponding author. Email: x.liu-22@tudelft.nl}}
  \address{Faculty of Electrical Engineering, Mathematics and Computer Science, Delft University of Technology, P.O.\ Box 5031, 2600 GA Delft, The Netherlands}
  \and
  \name{Piet Van Mieghem}
  \address{Faculty of Electrical Engineering, Mathematics and Computer Science, Delft University of Technology, P.O.\ Box 5031, 2600 GA Delft, The Netherlands\\
           Email: P.F.A.VanMieghem@tudelft.nl}
}

\maketitle

\begin{abstract}
{Our paper \cite{VanMieghem2025} described the stochastic approximation $\overline{rel}_G(p)=\bigl[1-\phi_D(1-p)\bigr]^{N}$ in \cite[eq. (2.2)]{VanMieghem2025} and the first-order approximation $(R_1)_G(p)=\prod_{i=1}^{N}\!\bigl[1-(1-p)^{d_i}\bigr]$ in \cite[eq. (4.1)]{VanMieghem2025} as upper bounds for the all-terminal reliability polynomial \(rel_G(p)\).
The present corrigendum clarifies that the unique upper bound is \(\Pr[\hat D_{\min}\geq 1]\), which is difficult to compute exactly, because we must account for correlated node-isolation events.
Both the stochastic approximation $\overline{rel}_G$ and the first-order approximation $(R_1)_G$ ignore those correlations, assume independence and, consequently, do not always upperbound \(rel_G(p)\) as stated previously. The complete graph \(K_{3}\) is a counterexample, where both approximations lie below the exact reliability polynomial $rel_{K_3}(p)$, illustrating that they are not upper bounds. Moreover, as claimed in \cite{VanMieghem2025}, the first-order approximation $(R_1)_G$ is not always more accurate than the stochastic approximation $\overline{rel}_G$. We show by an example that the relative accuracy of the stochastic approximation $\overline{rel}_G$ and the first-order approximation $(R_1)_G$  varies with the graph $G$ and the link operational probability $p$.
}{network robustness, node failure, probabilistic graph, reliability polynomial}
\end{abstract}

\section{Brief summary of our original statement}

In the original paper \cite{VanMieghem2025}, we investigated two \emph{degree-based} formulas for the all-terminal reliability polynomial
\(\mathrm{rel}_G(p)\): (i) our \emph{stochastic approximation} $\overline{rel}_G$  and (ii) the \emph{first-order approximation} $(R_1)_G$  due to
Jason Brown {\em et al.} in \cite{brown2021network}. We have argued that both \emph{degree-based} approximations are upper bounds for \(\mathrm{rel}_G(p)\).
Here, we first recall the relation that underlies these formulas and then explain the arguments in \cite{VanMieghem2025}.

Let \(\hat G\) denote the companion random graph, obtained by retaining each link of \(G\) independently with
probability \(p\), so that the reliability polynomial is \(\mathrm{rel}_G(p)=\Pr[\hat G\text{ connected}]\).
For any simple graph on \(N\ge 2\) nodes, connectivity of a graph implies the absence of isolated nodes, which means that the minimum degree $D_{\min}$  in the graph should be at larger than 1.  The opposite implication is
not always true, because a network can consist of separate, disconnected clusters containing nodes
each with minimum degree larger than 1. The event
\(\{\hat D_{\min}\ge 1\}\) should hold in the graph $\hat G$  to be connected, which leads to the equivalence
\[
   \bigl\{\hat G \text{ connected}\bigr\}
   \;\subseteq\;
   \bigl\{\hat D_{\min}\ge 1\bigr\}
   \qquad\Rightarrow\qquad
   \mathrm{rel}_G(p) \leq \Pr\!\bigl[\hat D_{\min}\ge 1\bigr]
\]

\begin{figure}[H]
  \centering
  \includegraphics[scale=0.4]{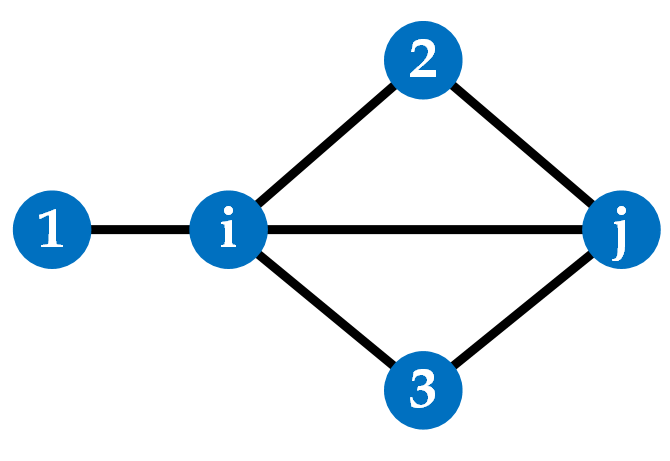}
  \caption{Illustration of dependence between events \{Node $i$ is isolated\} and \{Node $j$ is isolated\}. Each link is independently operational with probability $p$. The probability of event \{Node $i$ is isolated\} is $\Pr[\text{Node $i$ is isolated}]=1-p^4$. For node $j$, the probability is $\Pr[\text{Node $j$ is isolated}]=1-p^3$. The joint probability is $\Pr[\{\text{Node $i$ is isolated}\}\cap \{\text{Node $j$ is isolated}\}]=1-p^6$. If the two events were independent, the probability is $\Pr[\{\text{Node $i$ is isolated}\}\cap \{\text{Node $j$ is isolated}\}]=\Pr[\{\text{Node $i$ is isolated}\}]\Pr[\{\text{Node $j$ is isolated}\}]=1-p^3-p^4+p^7$.}
\label{fig:ind}
\end{figure}

In general, the computation of the probability \(\Pr[\hat D_{\min}\ge 1]\) is difficult. Due to the existence of common links, the node-isolation events are correlated, as exemplified in Fig \ref{fig:ind}. Both degree-based formulas assume independence and should thus be
viewed as approximations of \(\Pr[\hat D_{\min}\ge 1]\):

\begin{itemize}
  \item \textbf{The stochastic approximation in \cite{VanMieghem2025}.}
  Let \(\varphi_D(z)=\mathbb{E}[z^D]\) be the probability generating function of the degree $D$ of a node in the graph \(G\).
  Approximating the possible $N$ node isolation events by independent events gives
  \[
     \overline{\mathrm{rel}}_G(p)
     \;=\;
     \Bigl(1-\varphi_D(1-p)\Bigr)^{N}
     \;\approx\;
     \Pr\!\bigl[\hat D_{\min}\ge 1\bigr].
  \]

  \item \textbf{The first-order approximation (\cite{brown2021network}).}
  If $d_i$ denotes the degree of node \(i\), then
  \[
     (R_1)_G(p)
     \;=\;
     \prod_{i=1}^N \bigl(1-(1-p)^{d_i}\bigr)
     \;=\;
     \prod_{i=1}^N \Pr[\hat D_i\ge 1]
     \;\approx\;
     \Pr\!\bigl[\hat D_{\min}\ge 1\bigr],
  \]
  with equality to \(\Pr[\bigcap_i\{\hat D_i\ge 1\}]\) only under \emph{mutual independence} of the events
  \(\{\hat D_i\ge 1\}\).
\end{itemize}
Among these two degree-based approximations, the inequality
\[
  (R_1)_G(p)\;\le\;\overline{\mathrm{rel}}_G(p),
\]
 with equality only in regular graphs, was established in our original paper
(see \cite[Sec. 4.1]{VanMieghem2025}).

\paragraph{Clarification.}
In \cite{VanMieghem2025}, we have implicitly assumed that good approximations of the upper bound $\Pr\left[\hat D_{\min}\ge 1\right]$ also upperbound the reliability polynomial $\mathrm{rel}_G(p)$. Here, we show that this implicit assumption is not always correct. In addition, in the original paper \cite{VanMieghem2025}, we have claimed that $(R_1)_G(p)$ is always more accurate than $\overline{\mathrm{rel}}_G(p)$, which is also not generally true. Since both the stochastic approximation $ \overline{\mathrm{rel}}_G(p)$ and the first-order approximation $(R_1)_G(p)$  are independence-based and degree-based approximations of the upper bound $\Pr[\widehat D_{\min}\!\ge 1]$, either approximation can be closer to the reliability polynomial $\mathrm{rel}_G(p)$ depending on the graph $G$ and the operational link probability $p$, as demonstrated by a counterexample in Fig. \ref{fig:2upper_bound_with_inset} below.

\section{Corrigendum}

The inclusion of the events
\(
\{\hat G\ \mathrm{connected}\}\subseteq\{\hat D_{\min}\!\ge 1\}
\)
implies that the corresponding probabilities of the events obey
\[
  \mathrm{rel}_G(p)\;\le\;\Pr[\hat D_{\min}\!\ge 1].
\]
Both degree-based approximations $ \overline{\mathrm{rel}}_G(p)$ and $(R_1)_G(p)$ approximate the \emph{joint} event
\(\Pr[\bigcap_i\{\hat D_i\!\ge 1\}]\) by assuming independence of the
events \(\{\hat D_i\!\ge 1\}\) for each node $i$. The events \(\{\hat D_i\!\ge 1\}\) are dependent as follows from the basic law of the degree, 
stating that $\sum_{i=1}^N \hat D_i= 2L$, where $L$ is the number of links in the graph $G$.
In general, the joint probability $\Pr[\bigcap_i\{\hat D_i\!\ge 1\}]$ depends upon the graph's degree correlation structure, which is related to the assortativity \cite{PVM_review_assortativity} of the graph $G$.

In summary, there is no universal inequality for the degree-based approximations: neither lower bound 
$\overline{\mathrm{rel}}_G(p)\le \Pr[\widehat D_{\min}\!\ge 1]$ 
nor upper bound
$\overline{\mathrm{rel}}_G(p)\ge \Pr[\widehat D_{\min}\!\ge 1]$ 
holds for all graphs $G$ and all $p\in(0,1)$. A similar property applies to the first-order approximation $(R_1)_G(p)$. 
Depending on the graph $G$ and the probability $p\in(0,1)$, the stochastic approximation $\overline{\mathrm{rel}}_G(p)$ can satisfy either $\overline{\mathrm{rel}}_G(p) > \Pr[\widehat D_{\min}\!\ge 1]$ or $\overline{\mathrm{rel}}_G(p) < \Pr[\widehat D_{\min}\!\ge 1]$.
Consequently, neither degree-based
approximation is always an upper bound for \(\mathrm{rel}_G(p)\).

\section{Illustrative counter examples}
\subsection{The complete‐graph $K_{3}$}

In the complete graph $K_3$ on three nodes, every node has degree
$d_i=2$.  In regular graphs, the stochastic approximation
(Eq.\,2.2) and the first–order approximation (Eq.\,4.1) are the same as shown in \cite{VanMieghem2025} and both yield for $K_3$
\[
   \overline{rel}_{K_{3}}(p) \;=\; (R_1)_{K_{3}}(p)
   \;=\;\bigl[1-(1-p)^{2}\bigr]^{3}\;=\;(2p-p^{2})^{3}.
\]
However, the exact reliability polynomial in \cite{VanMieghem2025} for $K_3$, with the number of nodes $N=3$ and the number of links $L=3$, is
\[
\mathrm{rel}_{K_3}(p)=\sum_{j=0}^{1} F_j(1-p)^j p^{3-j}.
\]
where $F_j$ counts the sets of $j$ links whose removal leaves $G$ connected.
Clearly, if no node is removed, then $F_0=1$ and $F_1=3$, because deleting any one link of $K_3$
leaves a 3-node path, which is connected. After substituting these values of $F_j$, we obtain
\[
\mathrm{rel}_{K_3}(p)=1\cdot p^3 + 3\cdot(1-p)p^2
=3p^2-2p^3.
\]

If we define the difference
\(
  V(p)=(2p-p^{2})^{3}-(3p^{2}-2p^{3}),
\)
then a straightforward expansion shows
\(
  V(p)=p^{2}\!\bigl(-3+10p-12p^{2}+6p^{3}-p^{4}\bigr)<0
\)
for every \(0<p<1\) and \(V(0)=V(1)=0\).  Hence, it holds that
\[
   (2p-p^{2})^{3} \;<\; 3p^{2}-2p^{3}
   \quad\text{for }0<p<1,
\]
or, equivalently,
\[
   \overline{rel}_{K_{3}}(p)=(R_1)_{K_{3}}(p)<rel_{K_{3}}(p).
\]

This example of $K_3$ shows that both the stochastic approximation $\overline{rel}_{G}\left(
p\right)$ and the first-order approximation $\left(  R_{1}\right)  _{G}(p)$ {\em lower} bound and thus {\em not upper}  bound the reliability polynomial $rel_{G}\left(  p\right)$.

\subsection{The modified circulant on $N=15$ nodes}

\begin{figure}[H]
  \centering
  \includegraphics[scale=0.4]{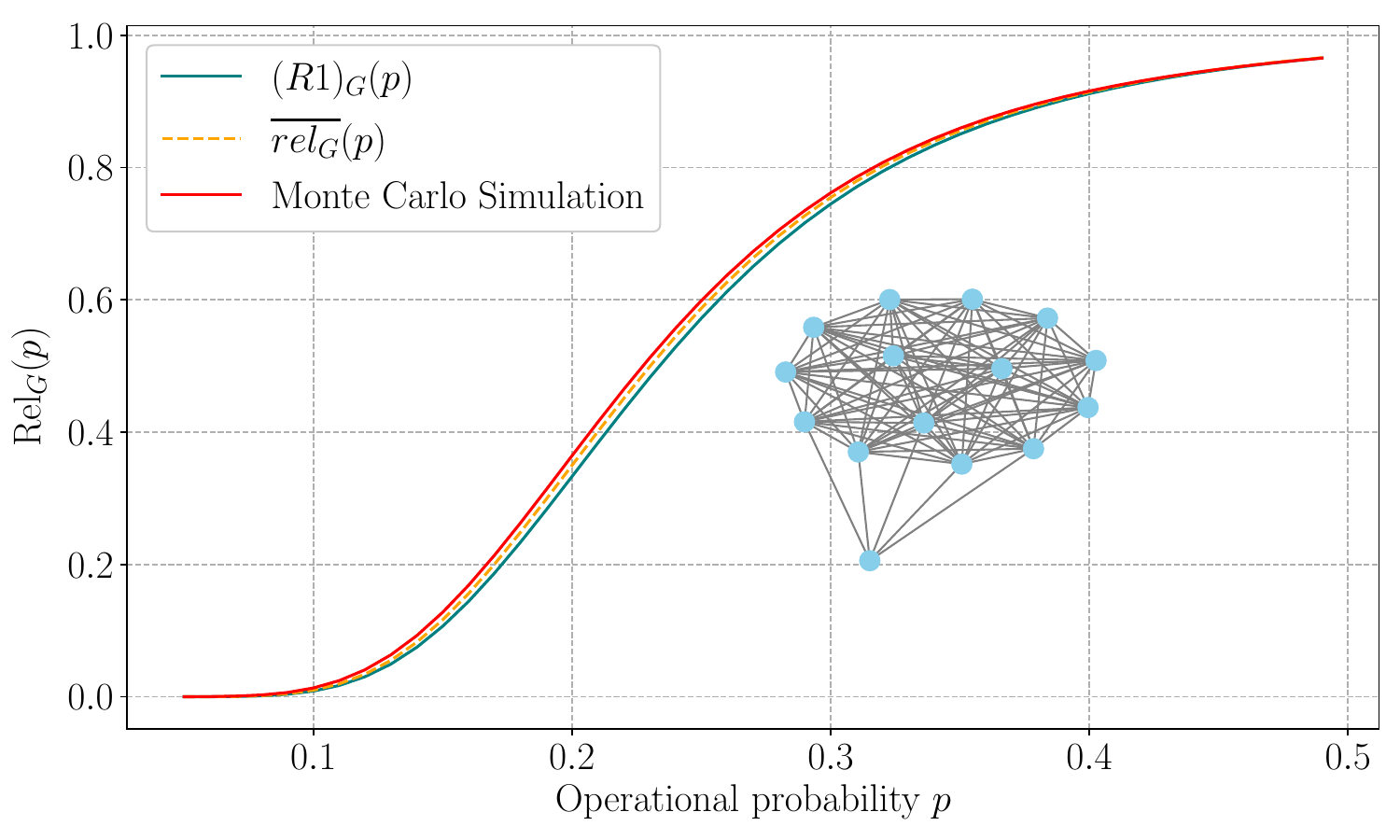}
  \caption{The stochastic approximation, first-order approximation and Monte Carlo simulations for a modified circulant on 15 nodes (nodes numbered \(1\)–\(15\)): start from the complete graph $K_{15}$ and delete the nine links \(1\!-\!2,\,1\!-\!3,\,1\!-\!4,\,1\!-\!6,\,1\!-\!7,\,1\!-\!8,\,1\!-\!13,\,1\!-\!14,\,1\!-\!15\). In the resulting graph, node $1$ has a degree of 5 (node \(1\) is linked only to nodes \(5,9,10,11,12\)), while all other nodes have degree \(13\) or \(14\). Properties of circulant matrices of small-world graphs are deduced in \cite[p. 194-200]{PVM_graphspectra_second_edition}.}
\label{fig:2upper_bound_with_inset}
\end{figure}

In the original paper \cite{VanMieghem2025},  we also claimed that $(R_1)_G(p)$ is always more accurate than $\overline{\mathrm{rel}}_G(p)$. This claim is not generally true, since there is no universal ordering between $(R_1)_G$, $\overline{\mathrm{rel}}_G$ and $\mathrm{rel}_G(p)$ among all graphs $G$ and link operational probability $p$.

Consider the modified circulant graph in Fig.~\ref{fig:2upper_bound_with_inset}, where node~1 is \emph{partially disconnected}, i.e., we delete $r\ge 1$ links incident to node~1 from the original circulant matrix, so that the degree of node 1 decreases by $r$, but node~1 remains non-isolated. Fig.~\ref{fig:2upper_bound_with_inset} shows, over a broad intermediate range of the link operational probability $p$, that
the stochastic approximation $\overline{\mathrm{rel}}_G(p)$ (dashed) stays closer to the Monte Carlo evaluation of $\mathrm{rel}_G(p)$ (red), which is very accurate and here regarded as benchmark,
than the first-order product $(R_1)_G(p)$ (solid). In summary, in this
example,
\[
  \bigl|\overline{\mathrm{rel}}_G(p)-\mathrm{rel}_G(p)\bigr|
  \;<\;
  \bigl|(R_1)_G(p)-\mathrm{rel}_G(p)\bigr|
\]
which contradicts the claim that $(R_1)_G$ is always more accurate than the stochastic approximation.


\end{document}